\documentclass[10pt, aps, nofootinbib]{revtex4}
\usepackage[latin1]{inputenc}
\usepackage[T1]{fontenc}
\usepackage{lmodern}
\usepackage{eurosym}
\usepackage[frenchb, english]{babel}
\usepackage{amsmath}
\usepackage{amsthm}
\usepackage{amssymb}
\usepackage{varioref}
\usepackage{listings}
\usepackage{graphicx}
\usepackage{float}
\usepackage{color}
\usepackage[usenames,dvipsnames]{xcolor}
\usepackage{epstopdf}
\usepackage{dsfont,units,pstricks}

\labelformat{Thm}{Theorem~#1}

\newcommand \be  {\begin{equation}}
\newcommand \beno  {\begin{equation*}}
\newcommand \bea {\begin{eqnarray} \nonumber }
\newcommand \ee  {\end{equation}}
\newcommand \eeno  {\end{equation*}}
\newcommand \eea {\end{eqnarray}}

\begin{document}

\title{On growth-optimal tax rates and the issue of wealth inequalities}

\author{Jean-Philippe Bouchaud}
\affiliation{CFM, 23 rue de l'Universit\'e, 75007 Paris, France, and Ecole Polytechnique, 91120 Palaiseau, France.}

\begin{abstract} 
We consider a highly stylized, yet non trivial model of the economy, with a public and private sector coupled through a wealth tax and a 
redistribution policy. The model can be fully solved analytically, and allows one to address the question of optimal taxation and of wealth inequalities. 
We find that according to the assumption made on the relative performance of public and private sectors, three situations are possible. Not surprisingly, 
the optimal wealth tax rate is either $0 \%$ for a deeply dysfunctional government and/or highly productive private sector or $100 \%$ 
for a highly efficient public sector and/or debilitated/risk averse private investors. If the gap between the public/private performance is moderate, 
there is an optimal positive wealth tax rate maximizing economic growth, even -- counter-intuitively -- when the private sector generates more growth. The compromise between profitable private investments and taxation however leads to a residual level of inequalities.  The mechanism leading to an optimal growth rate 
is related the well-known explore/exploit trade-off.
\end{abstract}

\sloppy
\maketitle

One of the most disgruntling aspects of the political debate is that intelligent people can defend in good faith totally 
conflicting points of view. This rarely happens in hard sciences, where either pure logic (as in mathematics) or empirical/experimental
data (as in natural sciences) eventually disqualify flawed arguments. While pure logic seems of little relevance when it comes to politics, 
identifying questions with possible empirical answers could perhaps help. 

As a rough caricature, the Left (or liberals) promotes big governments and correspondingly high taxes, when the Right (or conservatives) favours individual 
initiatives and private investments, and thus low tax rates. Both sides, one presumes, are concerned by the economic
welfare of their fellow citizens. One would like to set up a modelling framework where the question of optimal taxation 
can be investigated theoretically, with, for instance, long-term economic growth as a primary objective. This question is of course important enough
in itself; it has however received intense scrutiny in the wake of the recent debate about inequalities and their potential remedies,
introducing a global wealth tax for example \cite{Piketty}. If such a global tax was adopted, what should the taxation rate be? Are there theoretical guidelines
helping understand if there is a tax rate that, for example, maximizes economic growth? Surprisingly, most classical results in the economics 
literature conclude that the optimal wealth tax rate is zero, but the assumptions made are, as often, quite extreme \cite{AS,Judd,Chamley} and, for a recent
discussion, Ref. \cite{PS}. More recently, the problem of an optimal wealth tax rate has been revisited by Piketty and Saez \cite{PS}. They conclude that the wealth tax rate should be non zero, and actually 
quite high (at least in the form of an inheritance tax). However, their result hinges on the fact that the ``social welfare function'' that the government 
should optimize puts a strong weight on {\it those who receive zero bequest, and who must rely entirely on their labour income.} (from \cite{PS}). Also, the natural
fluctuations of economic growth (which is highly relevant in a multiplicative process, see below) and their impact on wealth inequalities are not 
addressed in their model, in which the main source of inequality comes the heterogeneous ``tastes'' of the previous generation of agents (some of which 
preferring not to bequeath anything to their children) and not, as in the model below, from successful investments.

The issue at stake here is of course related to whether inequality is promoting or maiming economic growth, and to the positive or negative 
role of redistribution \cite{Okun}. In a very interesting recent paper, the authors of \cite{IMF} provide both an insightful review of the literature and a thought-provoking
analysis of the best available data. Their broad conclusions are that a) more inequalities seem to hobble economic growth and b) moderate redistribution 
appears not to harm growth, up to a certain level beyond which detrimental effects appear. The aim of this note is to investigate these problems, in particular that of growth-optimal tax rates within the context of an admittedly\footnote{But note that our model is
no more stylized than those of the literature, including that of
Piketty and Saez \cite{PS}.} highly stylized, stochastic 
model of the economic activity, introduced in \cite{BM} and discussed in several subsequent papers, see e.g. \cite{Diego,Medo,Jap1,Jap2,Degond,Smerlak,Moukarzel} 
(see also \cite{Lux,Patriarca,Abergel} for reviews on this general class of models). The model is simple enough that it can be solved analytically, 
yet rich enough that the question of optimal taxation is non-trivial and allows one to clearly elicit the quantitative issues at stake in the Left/Right endless
skirmishes. Interestingly, the quantification of wealth inequalities can be simultaneously addressed within the same framework. We find that according to
the assumption made on the relative performance of public vs. private investments, three situations are possible. Not surprisingly, if the gap
between the average growth rate resulting from public and private investments exceeds some thresholds determined below, the optimal tax rate is either 
$0 \%$ (for a deeply dysfunctional government and/or highly productive private sector) or $100 \%$ (for a highly efficient public sector and/or 
debilitated/risk averse private investors). But if the gap between the two is moderate, there exists a well-defined wealth tax rate that leads to 
an optimal economic growth, {\it even when the private sector generates more growth}. The compromise between profitable private investments and taxation leads to 
a residual, ``incompressible'' level of inequalities that is necessary -- if ethical issues, among others, are disregarded --  to ensure maximum (asymptotic) wealth, 
even for the poorest. The non-trivial result here is that
even if on average private investors perform more poorly than public agencies, fluctuations associated to risky but highly profitable projects benefit
from reasonable tax rates -- and vice-versa: if private investors are only moderately ``better'' than the public sector, a non-zero wealth tax is still optimal. 
The mechanism leading to an optimal growth rate is related to the well-known explore/exploit trade-off -- in fact, our results below are strongly inspired 
by the analysis carried out in \cite{PRL} in the context of a more general growth and redistribution model. This mechanism was also discussed very recently in
a somewhat similar context in \cite{Peters,Volnew}.

Our toy economy is made of individuals who trade with one another and invest in risky projects, and a government that levies a wealth tax on these individuals, which
is channelled to government agencies to invest in other, presumably less innovative, projects (but see \cite{Mazzucato} for a forceful defense of the ``enterpreneurial state''). A fraction $f$ of the national wealth is redistributed to individuals, 
through various channels (social subsidies, public goods, technology transfers from public to private sectors, etc.). 
Denoting as $w_i(t)$ the wealth of agent $i$ at time $t$, and as ${\cal W}(t)$ the wealth of the state (excluding the private sector), the general structure of the model 
(introduced in this context in \cite{BM}, following earlier related ideas by Angle \cite{Angle}) is:
\be\label{Eq-gen}
\frac{\partial w_i(t)}{\partial t} = \sum_{j \neq i} J_{ij} w_j(t) - \sum_{j \neq i} J_{ji} w_i(t) + \eta_i(t) w_i(t) - \phi w_i(t) + f \kappa_i {\cal W}(t)
\ee
The first two terms encode trades, with $J_{ij}$ describing the profit margin made by $i$ when it trades with $j$, times the fraction of the 
wealth of $j$ involved in the trade. The third term describes the 
growth (or decay) of the wealth $w_i$ due to investments in risky projects, with a random growth rate $\eta_i(t)$ that can be decomposed 
into an idiosyncratic part (uncorrelated from agent to agent, who may make widely different investments) and a common part, reflecting how the socio-economic, 
intellectual, legal, etc. conditions for growth affect the whole private sector. We will therefore choose $\eta_i$ to be Gaussian, 
centred around a mean $m$, with the following simple correlation structure:\footnote{In fact, a better model should include a jump component 
with a positive skewness, corresponding to technological breakthroughs. We leave the investigation of this effect to further studies.} 
\be
\label{eq:CorrOU}
\langle \eta_i(t_1) \eta_j(t_2) \rangle - m^2  = \left(\delta_{ij} (1-\rho) + (1-\delta_{ij}) \rho \right) \frac{s^2}{2 \tau} e^{-\frac{|t_1-t_2|}{\tau}},
\ee
where the coefficient $\rho \in [0,1]$ gives the amplitude of the {\it common} part of the noisy private sector growth rate; while $s^2$ measures the
total variance of the returns on private projects. We assume, for simplicity, that $m$, $s^2$ and the persistence time $\tau$ are 
the same for all agents and for both the common and idiosyncratic contributions. Reasonable numbers, inspired by the US industrial production index since 
1954 \cite{Sarte} or the US stock market itself, might be $m = 3\%$ annual with a rms of $s = 10\%$ and a global correlation $\rho \sim 0.2$, 
and a persistence time $\tau$ of a few years. 

The last two terms of Eq. \ref{Eq-gen} describe a proportional wealth tax, with rate $\phi$, and the redistribution policy of the government, 
ensuring that a fraction $f$ of ${\cal W}(t)$ is distributed with weight $\kappa_i$ between the $N$ agents (with $\sum_i \kappa_i \equiv 1$).
In the following, we will only consider a proportional wealth tax, but an additional income tax can also be treated within the same formalism. 
Summing over all agents, we get the following equation for the total, privately owned wealth 
$W(t) = \sum_{i=1}^N w_i(t)$:\footnote{We assume for simplicity that no wealth is created through exchange per se, but this could be included if need be.
However, because of the very mechanism discussed in \cite{PRL} and in the present paper, exchange, by spreading wealth around, generically increases the average 
growth rate of the private sector!}
\be
\frac{\partial W(t)}{\partial t} =  \sum_{i=1}^N \eta_i(t) w_i(t) - \phi W(t) + f {\cal W}(t).
\ee

Assuming that the number $N$ of individuals is very large, the first term in the right-hand side can be rewritten, using Novikov's theorem, as:
\be
\sum_{i=1}^N \eta_i(t) w_i(t) \longrightarrow N \langle \eta_i(t) w_i(t) \rangle \approx  \left[m  + \frac12 (1 - \rho) s^2\right] W(t) + \sqrt{\rho} s \xi(t) W(t),
\ee
where we have assumed that $(m + s^2)\tau \ll 1$. The common part of the noise $\xi(t)$ is normalized to have unit variance. 
Therefore the dynamical equation describing private wealth simply reads:\footnote{Note that since the noise has a finite correlation time, we use 
Stratonovich's rule in the stochastic differential equations.}
\be\label{Eq-private}
\frac{\partial W(t)}{\partial t} =  (\widetilde m - \phi) W(t) + f {\cal W}(t) +  \sqrt{\rho} s \xi(t) W(t), \qquad \widetilde m = m + \frac12 (1 - \rho) s^2.
\ee

Similarly, the national wealth (excluding private wealth) obeys the following equation: 
\be\label{Eq-state}
\frac{\partial {\cal W}(t)}{\partial t} =  (\widetilde \mu - f) {\cal W}(t) + \phi W(t) + \sigma \xi'(t) {\cal W}(t),
\ee
where $\widetilde \mu$ and $\sigma$ are the analogue of $\widetilde m$ and $s$ for the public sector, and $\xi'(t)$ a noise term chosen for simplicity 
to have the same correlation time $\tau$ as $\xi(t)$ itself. The correlation between $\xi$ and $\xi'$, which is surely positive in reality, will have no 
bearing on the question of optimal taxation and inequality, so we set it to zero. Eqs. (\ref{Eq-private}, \ref{Eq-state}) 
are the starting point of the following analysis. They describe two random multiplicative growth processes, 
coupled by transfer terms parameterized by $f$ and $\phi$. Although seemingly simple, this model is generic and leads to an unexpected result, which is the 
existence of non-trivial optimal transfer rates that lead to a maximum growth rate of our toy economy. 

Let us first consider the uncoupled situation, $f = \phi = 0$. In this case, obviously, the growth rate of the private sector is $\widetilde m$ and that of the 
public sector is $\widetilde \mu$. Left-wing enthusiasts assume that the government is intrinsically more efficient than individuals, making collective decisions aimed at the long-term interest of its constituents -- so $\widetilde \mu > \widetilde m$. Therefore, intuitively, wealth should then be maximally taxed so that it 
gets managed by government agencies rather than by whimsical individuals. Right-wing diehards, on the other hand, deem the State intrinsically dysfunctional and wasteful,
whereas individual decisions, with more ``skin in the game'', are supposedly better informed and economically efficient. The ``invisible hand'' should then lead to  
$\widetilde \mu < \widetilde m$ and private wealth should be little taxed, or not at all. At this point, let us note that the individual growth rate 
{\it variance} enters positively to the definition of $\widetilde m$, which means that more risky investments in fact {\it increase} the long-time average growth rate 
of the private sector (this effect would even be stronger had we considered non-Gaussian noises with positive skewness, as seems fit to describe technological 
innovations). It could thus very well be that the bare private growth rate $m$ is indeed smaller than the public sector growth rate $\mu$, but that because
individuals can afford to take more risk than collective government agencies, the inequality gets reversed when one considers {\it dressed} growth rates 
$\widetilde m, \widetilde \mu$. The positive role of individual, risky initiatives is often put forth as an argument for reducing taxes on the ``high-net-worth''
individuals. 

We do not want to take side on this debate which, we believe, should be resolved on the basis of neutral empirical studies (following e.g. \cite{IMF}), and 
could heavily depend on countries, cultures and epochs. Our central point below is to show that provided the difference between growth rates $|\widetilde m - \widetilde \mu|$ is smaller than some 
thresholds $\Theta_\pm$, there always exists an optimal wealth tax rate $\phi$ for a given value of $f$ (or an optimal redistribution rate $f$ for a 
given value of $\phi$) such that the overall growth rate of the economy is maximized. When the difference in growth rate exceeds $\Theta_\pm$, 
however, the optimal policy is either to ``soak the rich'' ($\phi \to \infty$) if $\widetilde \mu > \widetilde m + \Theta_-$, 
or not tax them at all ($\phi \to 0$) if $\widetilde m > \widetilde \mu + \Theta_+$.  

In order to obtain these results analytically, we first transform the above equations Eqs. (\ref{Eq-private}, \ref{Eq-state}) by introducing $h(t) := \ln W(t)$
and $H(t):= \ln {\cal W}(t)$. As indicated above, we also assume that the private and public random noises, $\xi$ and $\xi'$, are uncorrelated. 
Any common component would only affect the overall growth rate of the economy, independently of the transfer rates $f$ and $\phi$. 
We further introduce $\Delta(t) = h(t) - H(t)$ to obtain an autonomous dynamical equation for its evolution:
\be\label{Eq-delta}
\frac{\partial \Delta(t)}{\partial t} = f e^{-\Delta(t)} - \phi e^{\Delta(t)} + \delta + \Xi(t); \qquad \delta \equiv \widetilde m - \widetilde \mu + f -\phi,
\ee
where $\Xi(t)$ is a new noise of variance $\Sigma^2= \sigma^2 + \rho s^2$ with the same correlation time $\tau$. In the limit where either $f \tau$ or $\phi \tau$ is much smaller than unity, 
the stationary distribution $P(\Delta)$ generated by the above equation can be computed as if the correlation time of the noise $\tau$ was zero, in 
which case it is given by the Gibbs measure: 
\be
P(\Delta) = {\cal Z}^{-1} e^{-\frac{2U(\Delta)}{\Sigma^2}}, \qquad U(\Delta) := f e^{-\Delta} + \phi e^{\Delta} - \delta \Delta,
\ee
where ${\cal Z}$ ensures that $P(\Delta)$ is normalized. Since $\Delta$ ($=h-H$) admits a stationary distribution, it is clear that the long-time, asymptotic growth of 
$h(t)$ and $H(t)$ must be the same, as $\sim e^{gt}$, and either of them can be used to define the growth rate ${g}$ of the economy. We therefore obtain $g$ from, e.g.:
\be\label{Eq-growth}
{g} = \left\langle \frac{\partial h(t)}{\partial t} \right \rangle_\Delta = \widetilde m + f \langle e^{-\Delta} \rangle_\Delta - \phi,
\ee
where $\langle ... \rangle_\Delta$ denotes averaging over the stationary distribution $P(\Delta)$. 
The final ingredient needed to compute ${g}$ is the following 
integral:
\be
\int_0^{\infty} {\rm d}u \, u^{\nu - 1} \, e^{-\beta u - \gamma/u} = 2 \left(\frac{\gamma}{\beta}\right)^{\nu/2} \, K_\nu\left(2 \sqrt{\beta \gamma}\right),
\ee
where $K_\nu(.)$ is the standard Bessel function of the second kind. Here we identify $e^{\Delta} \to u$, $2\phi/\Sigma^2 \to \beta$, $2f/\Sigma^2 \to \gamma$ and 
$2 \delta/\Sigma^2 \to \nu$ to obtain:
\be
\langle e^{-\Delta} \rangle_\Delta = \sqrt{\frac{\phi}{f}} \, \, \frac{K_{\nu - 1}(4 \sqrt{f \phi}/\Sigma^2)}{K_{\nu}(4 \sqrt{f \phi}/\Sigma^2)}. 
\ee
When this expression is plugged into Eq. (\ref{Eq-growth}), we obtain a fully analytic prediction for the growth rate $g$ of our toy-economy as a function of the different
parameters, in particular the wealth-tax rate $\phi$, valid in the limit where $\phi \tau \ll 1$. (For, e.g. $\phi = 2.5 \%$ and $\tau = 4$ years, 
$\phi \tau = 0.1$ which can still be considered small.)

We now study $g$ as a function of $\phi$ in the situation where $\widetilde m > \widetilde \mu$, i.e. when individuals' investments are smart enough, 
or bold enough, to generate a growth rate $\widetilde m$ larger than the public sector growth rate $\widetilde \mu$. We fix the redistribution rate $f$ 
and expand $g$ in the limit $\phi \to 0$, where the results will hold {\it independently} of the value of the correlation time $\tau$. 
Using the asymptotic behaviour of $K_\nu(z)$ for small arguments, namely:\footnote{The special case $\nu=0$ leads to $K_0(z) \approx_{z \to 0} - \ln(z/2) - 0.577...$.}
\be\label{BesselK}
K_\nu(z) \approx \frac{\Gamma(\nu)}{2} \left(\frac{z}{2}\right)^{-\nu} + \frac{\Gamma(-\nu)}{2} \left(\frac{z}{2}\right)^{\nu}, \qquad \nu \neq 0, 
\ee
one finds, when $\sqrt{f \phi} \ll \Sigma^2$:
\be
\langle e^{-\Delta} \rangle_\Delta \approx \sqrt{\frac{\phi}{f}} \frac{\Gamma(|\nu-1|)}{\Gamma(|\nu|)} \left(\frac{2 \sqrt{f \phi}}{\Sigma^2}\right)^{|\nu|-|\nu-1|}. 
\ee
When $\widetilde m > \widetilde \mu$ and $\phi \to 0$, the parameter $\nu$ is necessarily positive. We however need to consider separately two cases:

\begin{itemize}

\item $0 < \nu < 1$, i.e. the difference  $\widetilde m - \widetilde \mu$ is smaller than $\Sigma^2/2 - f$ (assumed to be positive here). In this case, one finds:  
\be \label{growth-max}
{g} \approx_{\phi \to 0} \widetilde m - \phi + \frac{\Gamma(1-\nu)}{\Gamma(\nu)} (\Sigma^2/2)^{1-2\nu} f^\nu \phi^\nu, 
\ee
which is an {\it increasing function} of $\phi$ for small $\phi$. This means that although individuals are better than the government at generating wealth, 
it is still optimal to introduce a wealth tax! This counter-intuitive effect is the main result of the present paper and will be discussed in more details below.

\item $\nu > 1$, i.e. the difference  $\widetilde m - \widetilde \mu$ is larger than $\Sigma^2/2 - f$.  Now, the small $\phi$ expansion of $g$ reads:
\be
{g} \approx_{\phi \to 0} \widetilde m + \left(\frac{2f}{(\nu - 1) \Sigma^2} - 1\right) \phi = \widetilde m +  \left(\frac{\widetilde \mu - \widetilde m + \Sigma^2/2}
{\widetilde m - \widetilde \mu + f  - \Sigma^2/2}\right) \phi,
\ee
which shows that as long as $\widetilde m - \widetilde \mu < \Theta_+ := \Sigma^2/2$, the growth rate is still, for small $\phi$, 
an increasing function of the wealth taxation rate, since the term between parenthesis is then always positive.\footnote{The next term in the 
small $\phi$ expansion of $g$ is of order $\phi^{\min(\nu,2)}$ with a {\it negative} coefficient, therefore leading to an optimal taxation rate 
that smoothly vanishes when $\widetilde m - \widetilde \mu  \to \Theta_+$.} But when the difference 
$\widetilde m - \widetilde \mu$ exceeds a threshold that grows with the global volatility of the economy $\Sigma^2$ 
it becomes optimal not to impose any wealth tax on the private sector. It is only in the case of a very efficient private sector, or a dysfunctional public sector 
with poor performances that optimal tax rate becomes zero.

\end{itemize}

Up to now, we have focused on the limit where $\phi \to 0$. The other limit where $\phi, f$ become very large is quite easy to analyze, since in that 
case $\Delta(t) \approx \Delta^* + \epsilon(t)$, with $\epsilon(t) \ll 1$. The dynamical equation for $\Delta$ thus becomes linear, and can be integrated for
an arbitrary noise correlator. The final result is that in this limit the growth rate ${g}$ takes a simple form that could have been guessed {\it a priori}:
\be \label{growth-asym}
{g} \approx \frac{f}{f + \phi} \widetilde m + \frac{\phi}{f + \phi} \widetilde \mu,
\ee
i.e. a weighted average of the public and private growth rate. The above analytical results suggest -- and this is confirmed by numerical simulations \cite{PSC} -- 
that when $\widetilde \mu < \widetilde m < \widetilde \mu + \Theta_+$, the growth 
rate ${g}(\phi)$ first grows, reaches a maximum for an optimal tax rate $\phi^*$, and then decreases to an asymptotic value given by Eq. (\ref{growth-asym})
that is always {\it below} the tax-free growth rate ${g}(\phi=0)=\widetilde m$ in the case considered here. The value of $\phi^*$ can be obtained from Eq. (\ref{growth-max}), provided it 
is found to be small (actually provided $2\sqrt{f\phi^*} \ll \Sigma^2$). Setting to zero the derivative of ${g}(\phi)$ with respect to $\phi$, one finds:
\be
(2\phi^{*}/\Sigma^2)^{1-\nu} = \frac{\nu \Gamma(1-\nu)}{\Gamma(\nu)} (2f/\Sigma^2)^\nu, 
\ee
for which the optimal growth rate ${g}^*$ exceeds the private sector growth rate $\widetilde m$ by an amount $(\nu^{-1}-1) \phi^*$. These formulae are only 
asymptotic and require $2f/\Sigma^2$ to be very small, and $\nu$ sufficiently far from $0$ and $1$, otherwise the sub-leading correction in Eq. (\ref{BesselK})
becomes relevant. The special case $\nu = 1/2$ is interesting since in this case $K_{\nu-1}(z) \equiv K_{\nu}(z)$ for all $z$. Therefore, 
the average growth rate becomes very simple and reads:
\be \label{growth-onehalf}
{g}= \widetilde m - \phi + \sqrt{f\phi}, \qquad (\nu = 1/2), 
\ee
which is maximized for $\phi^* = f/4$, for which the optimal growth rate is increased by an amount $f/4$ above the private growth rate $\widetilde m$, again, 
even when the government-induced growth rate is smaller than $\widetilde m$. Still, the global economy is better off with a wealth tax. The behaviour of the optimal taxation rate $\phi^*$ as a function of 
$\widetilde m -\widetilde \mu$ is sketched in Fig. 1.

\begin{figure}[t]
\includegraphics[width=8cm, height=6cm,clip=true]{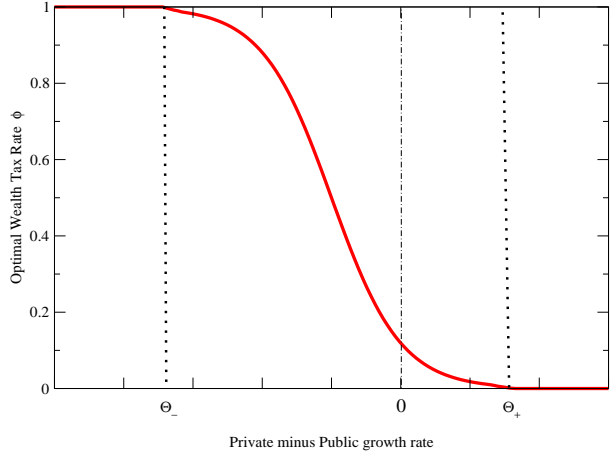}
\caption{\small Sketch of the dependence of the optimal tax rate $\phi^*$ as a function of the private-/government- induced growth rates 
$\widetilde m -\widetilde \mu$. The limit above which the wealth tax rate should be zero is $\Theta_+=\Sigma^2/2$, i.e. the larger the growth rate
volatility of either the private or the public sector (or both), the larger the region where a non-zero wealth tax is favourable.}
\label{fig1}
\end{figure}

We now turn to the case where private investments are less efficient than that of the public sector, either because the government promotes 
outstanding innovation through (risky) public research \cite{Mazzucato}, or because individuals are myopic (and invest in poor projects) 
or risk-averse (and hold their cash in non-productive assets), 
reducing $\widetilde m$ below $\widetilde \mu - f$. In the case $-1 < \nu < 0$, one obtains:
\be
{g} \approx_{\phi \to 0} \widetilde \mu - f  + \frac{\Gamma(1 - |\nu|)}{\Gamma(|\nu|)} (\Sigma^2/2)^{1 - 2 |\nu|} f^{|\nu|} \phi^{|\nu|},
\ee
where we kept the sub-leading term to retain a dependence on $\phi$, which otherwise disappears to leading order. We therefore see that, as expected in this case, small tax 
rates increase the growth rate of the economy. Note however that the linear term in $\phi$ has disappeared, which means that the optimal tax rate is not
in the small $\phi$ region anymore. The existence of an optimal taxation rate can be decided by analyzing the $\phi \to \infty$ limit, 
which is found to behave as:
\be
{g} \approx_{\phi \to \infty} \widetilde \mu + \frac{1}{\phi} \left[ \frac{\Sigma^2}{4\tau} - f (\widetilde \mu - \widetilde m - f) \right]. 
\ee
As expected, for infinite taxation rate, the economy is dominated by the public sector, and grows at rate $\widetilde \mu$. 
This asymptotic limit is however approached from below when $\widetilde \mu$ is large, i.e.  $\widetilde \mu
- \widetilde m > \Theta_- := f + {\Sigma^2}/(4\tau f)$ (in which case the theoretical optimal taxation rate is indeed $\phi = \infty$), and from {\it above} when 
$\widetilde \mu - \widetilde m < \Theta_-$, indicating the existence of a non-trivial optimal taxation rate $\phi^*$ that is no longer small in this case, 
but of order $\tau^{-1}$. 

In summary, one finds four regimes (see also Fig. 1), with $\Theta_+=\Sigma^2/2$ and $\Theta_-=f + {\Sigma^2}/(4\tau f)$:\footnote{The special case $\nu=0$, corresponding for example to a symmetric
situation $\widetilde \mu = \widetilde m$ and $\phi = f$, is interesting in itself, and leads to an optimal taxation rate very similar to what is found in \cite{PRL}.
Using the asymptotic behaviour of $K_0(z \to 0)$, one finds, to leading order ${g} \approx_{\phi \to 0} \widetilde m - \phi + \Sigma^2/2 \,
(\ln(\sqrt{f \phi}/\Sigma^2))^{-1}$, which is indeed a (strongly!) increasing function of $\phi$ at small $\phi$.} 

\begin{enumerate}

\item $\widetilde m > \widetilde \mu + \Theta_+$ (strong private sector/dysfunctional public sector): wealth should not be taxed at all, $\phi^* = 0$ 
(the right-wing/conservative lore). Note however that with no wealth tax (nor income tax), the private sector wealth would grow at a faster rate and asymptotically 
decouple from the national wealth, leading to an unsustainable situation similar to the `$r > g$' predicament discussed by Piketty \cite{Piketty}.

\item $\widetilde \mu < \widetilde m < \widetilde \mu + \Theta_+$ (private sector better than public sector but not by much): wealth should be taxed, all the more 
so when redistribution is stronger: $\phi^* \propto f^{\nu/1-\nu}$.

\item $\widetilde \mu - \Theta_- < \widetilde m < \widetilde \mu $ (public sector better than private sector but not much): wealth should be more strongly taxed, but 
some wealth should still remain in the private sector. The optimal tax rate is such that private wealth is preserved during the typical persistence scale $\tau$ 
of investment projects, i.e. $\phi^* \tau \sim 1$.

\item $\widetilde \mu > \widetilde m + \Theta_-$ (very strong public sector/dysfunctional private sector): wealth should be infinitely taxed for an optimal
growth rate of the economy (the communist hypothesis).

\end{enumerate}

Whereas the two extreme situations are quite expected (within our model), the intermediate two situations are not immediately intuitive; why, for example, should one tax the
rich at all when individuals are more efficient than government agencies, and how does this lead to an economic growth that is larger than that 
of the private sector itself? The mechanism is in fact deeply related to the general exploration/exploitation trade-off, along the lines of \cite{PRL,Peters}. 
The idea is simply that of {\it temporal diversification}: even though the
public sector might {\it on average} be less effective than individuals, some projects will be highly successful, and the government must have the means to finance
them, provided the generated growth efficiently redistributed in the private sector. Conversely, some private investments might be turn out to be bad and having 
under-allocated to them is a good idea on the long run. This is why the threshold $\Theta_+$ increases with the uncertainty of the investments: the difference between the performance
of the private and public sector must make up for the risk for a zero wealth tax rate to become optimal. The whole argument is of course reversed if the private sector was less effective than government agencies: it would still 
be beneficial to the economy to maintain a substantial level of private investment.

Let us finally discuss the issue of wealth inequalities within the population, which can be addressed within the same model \cite{BM,Smerlak}. 
As shown in \cite{BM,Jap1,Jap2}, Eq. (\ref{Eq-gen}) generically leads to a 
Pareto (power-law) tail for the distribution of wealth in the stationary state.\footnote{For a similar discussion within a discrete time model, see also \cite{Benhabib}. 
While not immediately obvious, their random multiplicative model is in fact deeply related to the model studied in \cite{BM}.} The tail exponent $\alpha$ can be computed explicitly in several cases \cite{BM,Jap1,Jap2}, 
and be associated to a Gini coefficient $G \approx (2 \alpha - 1)^{-1}$. The Gini coefficient is a standard measure of inequality \cite{Gini}; it is equal to zero for 
egalitarian societies, and to unity when a finite fraction of the total wealth is in the hands of a few individuals. The Gini coefficient of most advanced countries
is in the range $0.3 - 0.5$, corresponding to a Pareto tail index $\alpha$ in the range $1.5 - 2$. One expects that $\alpha$ should increase 
(and therefore the Gini coefficient decreases) as some wealth tax is introduced, as intuitively obvious. In the simple mean-field case where all the 
$J_{ij}$ in Eq. (\ref{Eq-gen}) are equal to $J_0/N$, one 
in fact finds that:\footnote{There is a subtlety here: one should take the limit $N \to \infty$ before the $t \to \infty$ limit to get a 
well behaved distribution of unscaled wealth, see \cite{Medo}. In fact the model can be rigourously mapped onto Derrida's Random Energy model, \cite{unpub}, and
can lead to wealth condensation for finite $N$.}
\be\label{eq-alpha}
\alpha(\phi) = 1 +  \frac{2(J_0+\phi)}{(1-\rho) s^2};\qquad \qquad \left[\Rightarrow G(\phi) \approx \frac{ (1-\rho) s^2}{4(J_0+\phi) + (1-\rho) s^2}\right],
\ee
an equation that remains approximately true when each agent trades with a large number of partners. (For a small number of trading partners, see Eq. (16) in Ref.
\cite{Jap2}.). Note that in the absence of tax, the wealth inequality is generated by the fact that agents have different (random) levels of success on their 
investments, parameterized by the idiosyncratic variance $(1-\rho) s^2$. The larger the variance, the larger the Gini coefficient (remember that the 
average growth rate $m$ is assumed to be the same for all agents -- systematic skill differences are therefore not accounted for and are not
responsible for creating inequalities here: randomness is enough!). The exchange intensity $J_0$, on the other hand, has the same redistribution effect 
as taxes on the Pareto index (or on the Gini coefficient), as originally noticed in \cite{BM}.

The conclusion of the analysis above is that depending on which of the four cases above described best the economy of a country, it may be optimal to levy
a non-zero wealth tax $\phi^*$, which would then act as to reduce inequalities {\it and} promote growth, as indeed suggested by the data \cite{IMF}. However, the resulting distribution of wealth will still be a highly skewed, Pareto
distribution, with a Gini coefficient $G^* \approx (2 \alpha(\phi^*) - 1)^{-1}$. Only in case 4. above (infinite tax rate) would it be optimal to remove
all inequalities from the society. But in all other cases, our model suggests that we should just come to terms with the fact that a certain level of wealth inequalities
is, in fact, optimal in terms of economic growth. In this respect, one should note that if the redistribution policy is fair (i.e. the coefficients
$\kappa_i$ are uniform over the population, or at least not concentrated on a small subset of agents), every member of the population benefits, at his own level, of
the optimal economic growth $g^*$ computed above. In more technical terms, this means that the whole distribution of wealth is scaled by the overall level of 
wealth $W(t)$. In this context, maximizing the overall growth rate appears to be optimal for everybody; however, other considerations might lead governments 
to reduce wealth inequalities further. Beyond any ethical considerations, strong inequalities come with an economic cost: social unrest and insecurity may indeed 
lead to a much reduced economic productivity (see the discussion in \cite{IMF}), i.e. smaller $\widetilde m$ that would in turn favour higher tax rates and 
steer the system towards lower inequalities. In other words, the Gini coefficient should enter on top of the growth rate the social welfare function that the government wants to 
optimize over both $\phi$ and $f$.\footnote{Possible non-linear effects, such as the negative dependence of $\widetilde m$ on $\phi$ 
(i.e. high taxes killing incentives for innovation \cite{Okun}) should also be taken into account.} One should however note that within our model, inequalities can in fact 
be reduced without increasing taxes, for example through more targeted redistribution policies (through the $\kappa_i$'s), or a better engineered 
exchange network, for example by increasing competition as to remove wealth ``gluts'' associated to nodes with locally convergent $J_{ij}$'s (i.e., the Apples of the world).

Although our model of the economy is obviously highly stylized (but not radically more than those in the economics literature), we believe that 
it helps uncover a non intuitive effect of taxation and redistribution that is a mere
consequence of the random multiplicative nature of economic growth. Our result on the existence of an optimal wealth tax rate even when individuals are 
better at generating growth than government agencies is non intuitive but follows from recent work on the existence of an exploration/exploitation trade-off in 
similar contexts \cite{PRL,Peters,Volnew}. We believe that it is, to some extent, a matter of detailed (and dispassionate) empirical investigation to determine in which of the above four
``phases'' the economy of a country most likely resides, before deciding on the level of wealth taxation (also accounting for the problem of tax evasion and/or 
brain drain, which would probably require a global wealth tax {\`a la} Piketty). 
In other words, whether $\widetilde m$ is larger or smaller than $\widetilde \mu$ cannot be decided on theoretical grounds, and requires special scrutiny. It may well be that individuals are more creative, or more risk takers, than public agencies, in which case a very small wealth tax would benefit all. Conversely, one may find out that rich individuals have a tendency to pile cash in their bank account, 
with little vision about where to invest optimally, or indulge in binge, unproductive consumption, therefore wasting resources and reducing growth opportunities
\cite{Mazzucato}. It should be the role of an efficient financial industry to help funnel savings into productive investments. We tend to believe that individuals, even wealthy, are not described by the rational {\it homo economicus} postulated by standard economic models, but rather prone to a slew of inefficient biases,\footnote{To paraphrase the usual saying: {\it If you are so rich, why aren't you smart?}, in particular when applied to inherited wealth.} well documented
in the behavioral literature \cite{Kahnemann,Wiki}. Smart, growth optimal investing is difficult and takes time; 
wealthy individuals may well need help and/or incentives to achieve it.\footnote{In this respect, differential tax rates could incentivize more growth-prone, 
risky investments and in turn increase $\widetilde m$.}

Models are useful to guide one's intuitions, but all the conclusions reached above should of course be taken with a grain of salt. We have insisted that 
attempting to determine some of the model's parameters is a prerequisite before drawing any serious policy recommendation. There are however
many effects left out of our framework, in particular the crucial question of time scales, that we have swept under the rug. As for most classical economics models, 
our results concern the long-run growth rate of the economy, where genuine multiplicative effects play a role. For growth rates and tax rates of a few percent per year, 
this means that we are looking at time scales of several decades before reaching the stationary state described above \cite{Gabaix}. On these long time scales, 
one should expect that the parameters of the model, in particular the growth rates themselves, will surely evolve, for instance
due to the feedback of the fiscal policy itself on the quality of the
infrastructures and on public and private incentives or the combination of other, political and societal factors \cite{Eric}. 
The impact of inequalities, education, corruption, etc. on these growth rates cannot be overstated \cite{IMF}. In spite of these interesting complications being left out, we hope that 
our model, based on minimal assumptions and a robust diversification argument, helps understanding why 
some amount of wealth tax might well be needed to increase long-term economic growth, and why some level of wealth inequality might 
be -- perhaps paradoxically -- socially beneficial.

Acknowledgements: This work was started as a very productive student project (``PSC'') at Ecole Polytechnique, with G. Bonnech\`ere, A. Carton, A. Gourdon, J. Guillaud, 
E. Pouille and Y. Zhang. Crucial discussions with Alexander Dobrinevski and Thomas Gueudr\'e on the exploration/exploitation trade-off, in particular the two-site 
problem, have been inspiring and led me to the idea of this paper. I also thank Jonathan Donier, Augustin Landier, Marc Potters, Matteo Smerlak and David Thesmar 
for very interesting insights on these issues. Finally, my gratitude to Giulio Biroli, Elisabeth Bouchaud, Ludovic Berthier, Catherine Fieschi, Marc M\'ezard (with whom the original model was developed), Ole Peters and Francesco Zamponi for very detailed and perceptive comments which helped improve the manuscript.



\end{document}